\documentclass[sts]{imsart}
\usepackage[svgnames]{xcolor}

\usepackage{listings}
\usepackage{color}

\definecolor{dkgreen}{rgb}{0,0.6,0}
\definecolor{gray}{rgb}{0.5,0.5,0.5}
\definecolor{mauve}{rgb}{0.58,0,0.82}

\usepackage{todonotes} %
\usepackage{url}

\usepackage{listings}

\RequirePackage{amsthm,amsmath,amsfonts,amssymb}
\newtheorem{theorem}{Theorem}
\RequirePackage[numbers]{natbib}
\usepackage{url}

\startlocaldefs

\endlocaldefs

\begin{document}

\begin{frontmatter}
\title{Distributionally robust and generalizable inference}
\runtitle{Replicability and generalization}

\begin{aug}
  \author[A]{\fnms{Dominik}~\snm{Rothenh\"ausler}\ead[label=e1]{rdominik@stanford.edu}}
  \and
\author[B]{\fnms{Peter}~\snm{B\"uhlmann}\ead[label=e2]{buhlmann@stat.math.ethz.ch}}
\address[A]{Dominik Rothenh\"ausler is Assistant Professor, Department of
  Statistics, Stanford University, United States of America \printead[presep={\ }]{e1}.}
\address[B]{Peter B\"uhlmann is Professor, Seminar for Statistics, ETH
  Z\"urich, Switzerland\printead[presep={\ }]{e2}.}
\end{aug}

\begin{abstract}
We discuss recently developed methods that quantify the stability and generalizability of statistical findings under distributional changes. In many practical problems, the data is not drawn i.i.d.\ from the target population. For example, unobserved sampling bias, batch effects, or unknown associations might inflate the variance compared to i.i.d.\ sampling. For reliable statistical inference, it is thus necessary to account for these types of variation. We discuss and review two methods that allow to quantify distribution stability based on a single dataset. The first method computes the sensitivity of a parameter under  worst-case distributional 
perturbations to understand which types of shift pose a threat to external validity. The second method treats distributional shifts as random which allows to assess average robustness (instead of worst-case). Based on a stability analysis of multiple estimators on a single dataset, it integrates both sampling and distributional uncertainty into a single confidence interval.
\end{abstract}


\end{frontmatter}

\section{Introduction}

Uncertainty quantification and inference in terms of confidence
statements in complex models has been a core topic in statistics over many
decades. In the last 10 years, substantial progress has been made for
high-dimensional and complex models, and we will briefly review these
developments in Section~\ref{sec:internal}. The main focus of this paper is different though, namely about 
generalizability and external validity of statistical findings and its
corresponding inference. In ordinary language: 
if a statistical result is significant in a study (i.e., a dataset), to
what extent can it be expected to be significant in another study which is
similar but not exactly of the same nature as the original one? This
question and corresponding solutions can be mathematically formalized, and
we will describe them in Sections~\ref{sec.external} - \ref{sec:calinf}. Such generalizability and external
validity of statistical inference is often of major interest in the context of empirical studies in e.g. medicine, public health or economics \cite{rothwell2005external,witteveen2020early}. 

To judge the generalizability and trustworthiness of a statistical result, it is crucial to investigate the fragility of the analysis. \citet{yu2020} discuss different types of perturbations that can be injected in the analysis process. If multiple datasets are available, one can adjust inference to account for the fact that the target population is different from the population at hand.
As an example, \cite{dahabreh2020} consider the problem of transporting inferences from multiple randomized trials to a new target population: the new population, which is not among the observed multiple datasets and potentially of slightly or moderately different nature is the one for which we want to generalize to. 

Having statistical inference tools which
are externally valid for somewhat different populations than the ones in
the data is a crucial component for improving replicability of statistical
(and scientific) results. The famous article by John Ioannidis
\cite{ioannidis2005} 
on the replicability crisis mentions major issues
about biases from reporting and different protocols. Distributionally
robust statistical procedures for confidence statements can be useful for
addressing an aspect of the replication problem, without explicitly aiming to
understand its possibly very diverse set of underlying reasons. 

\subsection{Internal and external validity}

We consider the setting where the data are realizations from either a
single data-generating distribution $P'$ or a set of data-generating
distributions $\{P'_e;\ e \in {\cal E}\}$ where $e$ is an index for a
sub-population and ${\cal E}$ is the space of observed sup-populations in
the data. We typically assume that the data are i.i.d. 
or
independent realizations depending on fixed covariates from these
distribution(s) $P'$ (or $P'_e$); but the framework also includes
sampling from a stochastic process or structured sampling in mixed effects
models.

An inferential statistical statement for a parameter $\theta(.)$ is called
\emph{internally valid} if it is 
statistically valid (or correct) for $\theta(P')$ or $\theta(P'_e)$ for some $e \in {\cal E}$.  Note that the parameter is a functional of the distribution $P$, $P_e$ or $P'$.
Thus, the parameter of interest $\theta(.)$ is a functional of a data generating
distribution from which the observed data arises. On the other hand,
\emph{external validity} is concerned about a parameter $\theta(P)$,
where $P \neq P'$ or $P \neq P'_e$ for all $e \in {\cal E}$. Thus,
external validity is about a parameter of a distribution which has not been
seen in the data, for example a regression parameter in new data which has
a different data generating distribution than the one generating the
observed (training) data.  

There is a fast growing literature on the theme of external validity,
including distributional
robustness \cite{sinha2017}, domain adaptation and transfer learning \cite{pan2010survey}, and transportability \cite{pearl2011transportability}.
We
will present a brief summarizing view of them in Section \ref{sec.external}. On the other
hand, there is very little work on distributionally robust confidence
statements. We review here some of the work from the latter topic \citep{gupta2021s,jeong2022calibrated} and provide an overarching perspective of the state-of-the-art.

\subsection{Internal sampling stability}\label{sec:internal}
Stability is an important concept to obtain higher degree of
replicability. The easiest version is internal sampling stability and is
often implemented via subsampling or bootstrapping the observed data
\cite{memepb09,mebu10,yu2013,buhlmann2014,huggins2022,yu2020}. Inspecting
and improving sampling stability is particularly useful for 
complex models and corresponding procedures: we mention here as some
examples
uncertainty assessment in high-dimensional models
\cite{zhangzhang11,vdgetal14,dezetal15,vanderpas2017,neykov2018}.  

Other forms of stability can be even used for external validity, and this
is discussed in Section~\ref{sec.external}.

\subsubsection{Post-selection inference.}
Since uncertainty quantification is difficult and often fragile in complex
models, post-selection inference procedures became rather popular
\cite{berk2013,lockhart2014,lee2016}. They are 
reliable and provide good internal replicability for the discovery of a particular selected hypothesis. However, if some data-driven model selection with e.g. the Lasso is performed \cite{lockhart2014,lee2016}, the entire procedure becomes often 
unstable and leads to a very bad degree of replicability. The reason
for it is as follows: the Lasso would typically pick a different set of
selected variables on another dataset (or a subsampled one) and hence, the
inference after selection will also focus on a different parameter and its
hypothesis: it is as much not replicable as the difference among the
selected models from the Lasso. This point is often not made very explicit
and things are expected to worsen when it comes to external validity.

\section{External validity of point estimation: distributional robustness, domain adaptation
  and causality}\label{sec.external}

External validity and corresponding (point) estimation strategies have been developed from different perspectives, all of them aiming to address the issue when the external (new) data has a different distribution than the original internal (training) data.
In the following, we give a high-level description of the topic. 

\subsection{Robust methods}
Protection against small-to-medium unknown perturbations can be achieved with robust methods. For large perturbations, these procedures become conservative. There is an important distinction between "classical" and distributional robustness. 

In the former "classical" case, the goal is to estimate a parameter of the unperturbed reference (or target) distribution when the (training) data is contaminated and often interpreted as realizations of a mixture of the reference and contamination distribution. There is only internal data, and the contaminations are among the observed samples. The methodology proceeds by data-driven down-weighting of outliers (contaminated data points), giving them less weight than $1/n$ with $n$ denoting the total (internal) sample size. See for example \cite{huber1964,hampeletal86}. 

In distributional robustness the aim is to predict well under adversarial perturbations in the external test data and the parameter of interest is with respect to a perturbed adversarial distribution. Here, the training dataset is internal and %
clean, while the contaminations or perturbations are not among the observed training data. This scenario is often 
relevant in modern machine learning. In this conceptual description, distributional robustness arises from up-weighting certain data points (giving them more weight than $1/n$) in order to achieve good performance on test data. For example, in regression one would aim to estimate a function $f(.)$ which optimizes
\begin{eqnarray*}
\mbox{argmin}_{f(.) \in {\cal F}} \sup_{P; d(P,P') \le \rho} \mathbb{E}_{P} [(Y - f(X))^2],
\end{eqnarray*}
where ${\cal F}$ is a suitable class of functions, $P'$ is the internal training distribution, $P$ is the external distribution, $d(.,.)$ a metric between probability distributions, $\rho$ a certain positive number, 
$Y$ a univariate response, and $X$ the vector of covariates (and $(Y,X) \sim P$) \cite{ben2002robust,bertsimas2011theory}.

\subsection{Domain adaptation and re-weighting}
Domain adaptation methods can cope with large distributional shift and perturbations from the training data distribution $P'$ to a target (test) distribution $P$ which generates new data. This can be achieved by re-weighting which takes the distributional change into account \cite{long2014transfer,li2016prediction}.

For example, one might be interested in
\begin{align*}
    &\mbox{argmin}_{f(.) \in {\cal F}} \mathbb{E}_{P} [(Y - f(X))^2] \\
    &= \mbox{argmin}_{f(.) \in {\cal F}} \mathbb{E}_{P'} [(Y - f(X))^2 w(X,Y)],
\end{align*}
Here, $w(X,Y) = \frac{dP}{dP'}(X,Y)$ is the Radon-Nikodym derivative. The results above motivate weighted empirical risk minimization:
\begin{equation*}
    \hat f = \mbox{argmin}_{f(.) \in {\cal F}} \frac{1}{n} \sum_{i=1}^n \hat w(X_i,Y_i) (Y_i - f(X_i))^2,
\end{equation*}
for some estimate $\hat w(\bullet)$ of $w(\bullet)$. There is often an assumption that restricts the shift in a particular way. For example, if $w(\bullet)$ only depends on $X$, we are in the popular setting of \emph{covariate shift} \cite[cf.]{quinonero2009dataset}. There is an underlying assumption about some overlap between the training and target (or test) distribution which then enables adapting to a different domain which may be far away in terms of a probabilistic distance.

In a different line of work, one tries to learn invariant representations of the features \citep{pan2010survey,baktashmotlagh2013unsupervised}. This is based on the idea that if a representation of the data is invariant between the training distribution $P'$ and target distribution $P$, then feeding these representations into a prediction algorithm might exhibit improved generalizability compared to feeding the untransformed data into a prediction algorithm. 

 The empirical success of such domain adaptation methods and algorithms is primarily documented in the field of computer vision \cite{gopalan2011domain,gong2012geodesic,peng2019moment}: indeed, it is remarkable that even though $d(P,P')$ is large for a metric $d(.,.)$, it is possible to accurately learn from $P'$ some aspects about $P$.  

\subsection{Invariance and causality} 
Another framework for achieving external validity is to learn causal representations which are able to generalize well outside the internal data. This includes invariance of feature representations 
\cite{heinze2021conditional}, or of residuals and learning some causal structures \cite{jonipb16,rojas2018,meinshausen2018,heinze2018,pb20,rothenhausler2021}. With such \emph{structural} approaches and models, no overlap assumption between $P'$ and $P$ is required but they typically rely on multiple sources or environments for learning the invariances. Unlike distributional robustness but in the same vein as domain adaptation, these methods allow for large distributional shifts and interventions between the internal and external data-generating distributions. 

\subsection{The role of multiple sources or environments}
Internal training data which is grouped according to different sources or groups under different environments, denoted above by $e \in {\cal E}$ with ${\cal E}$ being the space of observed environments, provides useful information for external generalization. The main reason is that internal sources of heterogeneity can be used to model distributional shifts, invariances or infer causal structure. Such multi-source/environment information has been exploited from a theory and practical point of view: for optimizing worst environment risk \cite{meinshausen2015maximin,buhlmann2015magging,sagawa2019distributionally}, for domain adaptation \cite{gong2012geodesic,baktashmotlagh2013unsupervised,heinze2021conditional,chen2021domain}, and for causal regularization aiming to obtain invariant residuals \cite{jonipb16,heinze2018,rothenhausler2021,arjovsky2019invariant}.

We also note that instrumental variables regression is related to multi-environment problems \cite{angristetal96,imbens2014,imru15}. If the instruments are discrete, they can be thought as encoding different environments, but IV regression also covers continuous forms of heterogeneity. A main and strong assumption is the so-called validity of such instruments: under such strong conditions, the invariant structure is equal to the causal structure, and a causal model is also externally valid under arbitrarily strong perturbations of the covariates.

\section{Distributionally robust uncertainty quantification}\label{sec:externalstability}

Considerations of external validity not only affect (point) estimation strategies, but should also affect how we report uncertainty. If data from multiple environments are available, one can conduct some type of meta-analysis. For example, partial conjunction tests \citep{heller2007conjunction,benjamini2008screening,wang2019admissibility}  allow to conduct valid inference in situations where a few of the datasets are perturbed. Such analysis provides internal validity among the different environments only. It cannot go beyond the internal multi-environment data. If only one dataset is available, there exist much fewer methods that account for distributional uncertainty. Existing methods either
\begin{itemize}
    \item employ worst-case bounds between the distribution of $P'$ and $P$; or
\item assume that the probabilities of events change randomly between $P'$ and $P$. 
\end{itemize}
As an example of the first approach, assume that we know that $\mathrm{D}_{KL}(P \| P') \le \delta$, where $\mathrm{D}_{KL}( . \| . )$ denotes the Kullback-Leibler divergence, 
and that $X_i \stackrel{\text{i.i.d.}}{\sim} P'$, with $P' = \mathcal{N}(\mu,1)$.  We aim to construct a confidence interval $I = I(X_1,\ldots,X_n,\delta)$ which is uniformly valid over the Kullback-Leibler ball, that is:
\begin{equation}\label{eq:unifci}
   \inf_{P: \mathrm{D}_{KL}(P \| P') \le \delta} \mathbb{P}[ \mathbb{E}_P[X] \in I] = 1-\alpha.
\end{equation}
Using some algebra, we get that
\begin{equation*}
   I = \frac{1}{n} \sum_{i=1}^n X_i  \pm \left( \frac{z_{1-\alpha/2}}{\sqrt{n}} + \sqrt{2 \delta} \right)
\end{equation*}
satisfies equation~\eqref{eq:unifci}, where $z_{1-\alpha/2}$ is the $1-\alpha/2$ quantile of a standard Gaussian random variable. This robust confidence interval is similar in spirit to robust versions of the probability ratio test \citep{huber1965robust} in the sense that one needs to pre-specify the strength of perturbations $\delta$.
Note that this confidence interval has a component that does not converge to zero as $n \rightarrow \infty$. 

Another approach is given by sensitivity analysis in causal inference, which investigates the stability of a statistical finding under (potential) unobserved confounding  \citep{cornfield1959smoking,rosenbaum1987sensitivity}. Today this is a field of active research \citep{ding2016sensitivity,zhao2019sensitivity,cinelli2020making,yadlowsky2018bounds,dorn2021doubly,jin2021sensitivity}. Such
sensitivity analysis is of the following nature.  
First, the tools are usually specific to the estimation strategy,
and thus have to be used on a case-by-case basis. Secondly, when considering worst-case distributional perturbations, very small shifts can already change results substantially. Since most sensitivity analyses are measuring worst-case stability, reported "instabilities" often occur due to the overly conservative worst-case analysis. 

In the following, we describe how these issues can potentially be addressed by a different type of sensitivity or stability analysis which is takes a ``directional worst-case'' view point.

 \subsection{Towards general-purpose tools for stability analysis}
 
 Often, practitioners are interested in \emph{sign stability} of a one-dimensional statistical parameter.  To be more specific, one might want to infer whether $\text{sign}(\theta(P)) = \text{sign}(\theta(P')) \approx \text{sign}(\hat{\theta})$
 for a reasonable set of perturbed distributions $P \in \mathcal{P}$. The motivation behind sign stability is that the parameter might correspond to whether or not a medication has a positive effect. 
 We can quantify the sign stability by estimating
 \begin{align}\label{eqdefs}
  s = &\exp(-\inf_{P}  \mathrm{D}_{KL}(P \| P')  \nonumber \\
    &\text{ such that }  \text{sign}(\theta(P')) 
   \neq \text{sign}(\theta(P))).
 \end{align}
 In example above, sign stability captures whether a medication that has a beneficial effect on the observed population might be harmful under distribution shift.

 In analogy to the $p$-value, if $s$ is close to zero, the sign of $\theta(.)$ is very stable under distributional changes. On the other hand, if $s$ is close to one, the sign is highly unstable under distributional changes.

 Let's consider an example. Assume we are interested in estimating the mean $\theta(P) = \mathbb{E}_P[X]$. Donsker and Varadhan \citep{donsker1976asymptotic} showed that if the moment generating function of $X$ is finite, then 
 \begin{equation*}
     s = \inf_\lambda \mathbb{E}_{P'}[e^{\lambda X}],
 \end{equation*}
 where $P'$ is from the data generating distribution and hence can be inferred from observed data.
 In fact, for i.i.d.\ observations $X_i \sim P'$ we can use the following plug-in estimator of the distributional stability measure $s$:
 \begin{equation*}
     \hat s = \inf_\lambda \frac{1}{n} \sum_{i=1}^n e^{\lambda X_i}
 \end{equation*}
Consistency guarantees for this estimator of stability are given in \cite{gupta2021s}. For other estimands than the expected value, such as parameters in generalized linear models or estimands defined via moment equations, estimating $s$ is more involved since Donsker-Varadhan does not apply directly. 

     In practice, one can use simple linear approximations to estimate $s$. If the estimand $\theta(.)$ is differentiable as a functional on the distribution space, by definition
\begin{equation}
    \theta(P) - \theta(P') = \mathbb{E}_P[\phi_{P'}(D)] + o(d_K(P,P'))
\end{equation} 
for some metric $d_K(.,.)$ such as the Kolmogorov metric and a function $\phi_{P'}(D)$ with $\mathbb{E}_{P'}[\phi_{P'}(D)] = 0$. 
Then Donsker-Varadhan suggests using the estimator
 \begin{equation}\label{eqhats}
     \hat s = \inf_\lambda \frac{1}{n} \sum_{i=1}^n e^{\lambda  (\hat \theta + \hat \phi(D_i))}.
 \end{equation}
 where $\hat \phi$ is an estimate of the influence function $\phi_{P'}(.)$ and $\hat \theta$ is an estimate of $\theta(P')$. For example if $(X,Y) \in \mathbb{R}^{p+1}$ 
 and $\theta(P)$ is the $k$-th component of the regression vector, that is $\theta(P) = \beta_k(P)$, where
 \begin{equation*}
     \beta(P) = \arg \min_\beta \mathbb{E}_P[(Y- X \beta)^2],
 \end{equation*}
 then one can estimate $\phi_{P'}(D_i)$ via $$ \hat \phi(D_i) =   ( \frac{1}{n} \sum_{j=1}^n  X_j^\intercal X_j)_{k,\bullet}^{-1}  X_{i}^\intercal (Y_i - X_i \hat \beta), $$
where $\hat \beta = \arg \min_\beta \frac{1}{n} \sum_{j=1}^n (Y_j- X_j \beta)^2$. As before, the data $(X_i,Y_i)_{i=1,\ldots,n}$ is drawn i.i.d.\ from $P'$. This strategy allows to estimate $s$ for common estimands such as parameters of generalized linear models or parameters defined via moment equations. Algorithms with consistency guarantees are given in \cite{gupta2021s}. 

These $s$-values can then be compared to benchmarks. As an example, in \cite{devauxquantifying}, the authors compute benchmarks for different national surveys (e.g., ANES and CES).  To be more concrete, they 
estimate benchmarks $\hat b_{P, P'} = e^{- \hat{\mathrm{D}}_{KL}(P \| P') }$ for $P$ and $P'$ corresponding to different national surveys. They
report that the empirical mean of the estimated $\hat b$, averaged over multiple pairs of surveys is $.86$.
This puts distributional stability measures into context. For example, if the $s$-value of a statistical result in a similar application is larger than $.86$, then this is an indication that the result might not generalize; in the sense that a distribution shift of the size that is observed between different national surveys is large enough to change the sign of the result. Note that for simplicity we have ignored statistical uncertainty quantification. Details on how to compute confidence intervals for $s$-values can be found in \cite{gupta2021s}. 

Of course, distributional stability measures are context dependent. In \cite{devauxquantifying} it is argued that one has to choose a smaller threshold when generalizing from national surveys to samples from Amazon Mechanical Turk (MTurk).

  Under arbitrary distribution shifts, one will usually be able to change the sign of statistical parameters under very small shifts. Thus, to make the tools more useful in practice, it is important to restrict the class of considered distribution shifts.
  
\subsubsection{Beyond omni-directional shifts}

In principle, any statistical finding breaks down under arbitrary distributional shifts. Thus, a practitioner might be interested in learning under what circumstances (that 
is, under which type of distribution shifts) a result breaks. As an example, maybe a correlation between two variables is nearly invariant across populations with different socioeconomic status, but highly variable across different age groups.  

In the following we will discuss how targeted sensitivity or stability analysis can be formalized. Related to the definition of $s$ in \eqref{eqdefs}, for a random variable $E$ which is observed in the data,
 \begin{align}\label{eqdefsE}
 \begin{split}
  s_E &= \exp(- \inf_{P: P[ . |E] = P'[. | E]}  \mathrm{D}_{KL}(P \| P') \\
  &\text{ such that } \text{sign}(\theta(P)) \neq \text{sign}(\theta(P')).
  \end{split}
 \end{align}
Formally, 
$s_E$ is a deterministic value that lies in $[0,1]$; the constraint $P[ . |E] = P'[. | E]$ is meant to hold almost surely w.r.t. $E$. 
  In words, we investigate how much a shift in the marginal distribution of $E$ can affect the parameter, while keeping the conditional distribution $P[.|E]$ invariant.

 Estimation of this stability parameter relies on a variation of Donsker-Varadhan's Lemma \citep{donsker1976asymptotic}. Using a similar approximation as for equation~\eqref{eqhats},
 $s_E$ can be estimated via
 \begin{equation}\label{eq:directional-approx}
    \hat{s}_E =  \inf_{\lambda} \frac{1}{n} \sum_{i=1}^n e^{\lambda (\hat \theta + \hat Q(E_i))}
 \end{equation}
 where $\hat Q(e)$ is an estimate of $Q(e) = \mathbb{E}_{P'}[\phi_{P'}(D)|E=e]$. A formal justification of this estimator is given in \cite{gupta2021s}.
 
 Let's return to the case of linear regression with $X=E$, with $X$ being potentially multi-dimensional. 
 This means we are considering a distribution shift in the covariates while keeping $Y|X$ constant. For $\theta(P) = \beta_k(P)$, a short calculation shows that
 \begin{align*}
 Q(X_i) = \mathbb{E}_{P'}[X^\intercal X]_{k,\bullet}^{-1} X_{i}^\intercal ( \mathbb{E}_{P'}[Y|X=X_i] - X_i \beta(P') ).
 \end{align*}
Based on this observation one can form a plug-in estimator of  $Q(x)$ and use equation~\eqref{eq:directional-approx} to estimate the directional stability coefficient $s_E$. Note that if the model is well-specified, $\mathbb{E}_{P'}[Y|X=x] - x \beta(P') = 0$ and thus the stability value $s_X $ is zero as long as $\theta(P') \neq 0$. Thus, for the specific choice of $E=X$, the directional stability coefficient $s_E$ captures whether the model is well-specified.

\subsubsection{Real-world example}

We demonstrate the usage of the stability measure $s_E$ in \eqref{eqdefsE} on the life-cycle savings data \cite{belsley1980regression}. The dataset contains measurements of the ratio between personal savings divided by disposable income (savings ratio - sr), the percentage of population under 15 (pop15), and the percentage of population over 75 (pop75), the disposable income (dpi) and the growth rate of the disposable income (ddpi). Under Modigliani's life-cycle savings hypothesis \cite{modigliani1966life}, the savings ratio is explained by these four covariates. We want to investigate the robustness of the linear model under various distributional shifts. A package implementing the estimation of the directional stability measure $s_E$ from \eqref{eqdefsE} as in  
\eqref{eq:directional-approx} 
can be obtained from \url{github.com/rothenhaeusler/stability}. The following R code fits a linear model and computes the stability of the linear regression coefficient corresponding to pop15, that is, the stability of $\theta(P) = \beta_{\mathrm{pop15}}(P)$ for different choices of $E$. We consider distribution shift both in single components of $X$, but also in the outcome $Y$:
\begin{verbatim}
> fit <- lm(sr ~ pop15 + pop75 + 
    dpi + ddpi, data = LifeCycleSavings)
> stability(fit,param="pop15")
Stability values
 
   s_sr s_pop15 s_pop75   s_dpi  s_ddpi 
  0.863   0.368   0.781   0.687   0.837
\end{verbatim}
Here, the names of the different columns correspond to the different choices of $E$. %

These values can be used to compare the relative stability of parameter values under different choices of $E$, for example using the smallest $\hat{s}_E$ as the baseline stability value. 
In addition, these robustness or stability measures can be compared to reference values computed across real-world datasets, as discussed in the previous section. For the population of the United States, based on the census of 2016, we get an estimate of the benchmark value $\hat b = .17$ for the change in the distribution of pop15. As $\hat b = .17$ is smaller than $\hat s_{pop15} = .368$ (see the output above), we have to be concerned that due to a large shift in the distribution of pop15, the sign of the regression coefficient might change if we were to collect new data from the US. On the other hand, for Mexico the estimate is $\hat b = .54$ which is larger than $\hat s_{pop15} = .368$. This suggests that we do not have to be concerned that a shift in pop15 changes the regression coefficient if we were to collect new data from Mexico. Such calculations allow us to gauge the extent to which a result will generalize. %

\begin{figure}
\begin{center}
    \includegraphics[scale=.4]{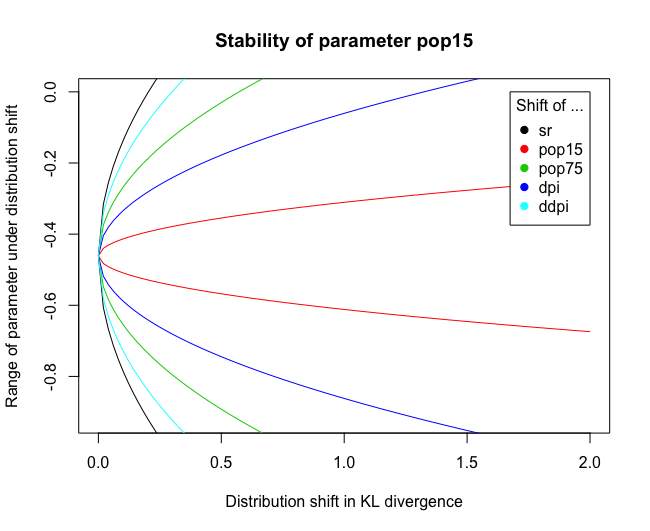}
\end{center}
\caption{Stability of the parameter pop15 under various distributional shifts as reported by \texttt{stability()}. Each coloured pair of lines corresponds to the different individual components of the parameter vector and displays the estimated versions of \eqref{eq:lowerbd} and \eqref{eq:upperbd}, respectively.
}\label{fig:stability}
\end{figure}

In addition to s-values, the \texttt{R}-function \texttt{stability} provides a visualization of the stability of parameters under distributional shifts. More concretely, for different choices of $E$ and an upper bound on the distribution shift $x$ we compute upper and lower bounds for parameter values as follows:
\begin{align}
  y_{\text{upper-bound}} & =  \sup \theta(P) \text{ such that }   \label{eq:lowerbd} \\
      & P'[ . |E] = P[ . | E] \text{ and } \mathrm{D}_{KL}(P \| P') \le x \nonumber  \\
  y_{\text{lower-bound}} &=  \inf \theta(P) \text{ such that }  \label{eq:upperbd} \\
  & P'[ . |E] = P[ . | E] \text{ and } \mathrm{D}_{KL}(P \| P') \le x  \nonumber
\end{align}
In Figure~\ref{fig:stability}, we visualize estimated versions (based on a linear approximation and plug-in)
of these upper and lower bounds across $x$ for different choices of the variable $E$.

This plot allows to derive bounds on parameters based on background knowledge. For example, if the data scientist expects that the distribution of dpi is expected to shift by at most $.5$ in Kullback-Leibler divergence between settings, then one would obtain an (estimated) upper bound of $-.2$ for the parameter.   

\section{Confidence intervals that account for both sampling and distributional uncertainty}\label{sec:calinf}

The diagnostic tools discussed in the previous section can be conservative since they still rely on possibly directional worst-case bounds. In practice, we need not be that pessimistic and thus we consider here \emph{average perturbation} effects. Furthermore, we will describe procedures which do not rely on the user's interpretation of stability values but \emph{estimate the amount of perturbations} from data.

For our further developments, we model the perturbation process as random. Intuitively speaking, if the data $D_1,\ldots,D_n$ is drawn i.i.d.\ from the sampling distribution $P'$ which randomly deviates from the target distribution $P$, we can decompose uncertainty into a sampling component and a distributional component:
\begin{equation*}
    \theta(P_n') - \theta(P) = \underbrace{\theta(P_n') - \theta(P')}_{\substack{\text{sampling uncertainty}}}  + \underbrace{\theta(P') - \theta(P)}_{\substack{\text{distributional uncertainty}}}.
\end{equation*}
Here, $P_n'$ denotes the empirical measure of $D_1,\ldots,D_n$.

Compared to classical statistical inference, 
we aim to construct confidence intervals for the target $\theta(P)$, instead of the parameter of the sampling distribution $\theta(P')$. However, without any restrictions on the perturbation process, it is impossible to quantify the magnitude of $\theta(P') - \theta(P)$. This raises the question about the 
distributional perturbation model.

\subsection{A model for distributional perturbations}\label{sec.distrpert}

An easy and perhaps 
natural distributional perturbation model is as follows.
For any event, the perturbed distribution will assign slightly different probabilities compared to the target distribution $P$. To simplify the discussion in the following we will assume that the sample space $\mathcal{D}$ is discrete with uniform weights on the singletons, that means for all $d \in \mathcal{D}$
\begin{equation*}
    P[D=d] = \frac{1}{|\mathcal{D}|}.
\end{equation*}
A perturbed distribution can now be formed by drawing exchangeable random variables $\xi_i \ge 0$ ($i=1,\ldots ,|\mathcal{D}|)$ with $\sum_i \xi_i =1$, and setting
\begin{equation*}
    P^\xi[D=d_i] = \xi_i.
\end{equation*}
Although the discussion in this section has focused on the discrete case, a similar argument works in the continuous case, see \cite[Sec.2]{jeong2022calibrated}.

\subsubsection{Mean and variance of sample means under the perturbation model.}
For simplicity, we consider first the sample mean. Conditionally on $\xi$, the data is drawn i.i.d.\ from $P^\xi$ and we denote the marginal distribution when averaging over the random variables $\xi_i\ (i=1,\ldots ,|\mathcal{D}|)$ by 
\begin{eqnarray*}
\mathrm{d}P_{\text{marginal}}(d) = \int \mathrm{d}P^{\xi_1,\ldots ,\xi_{|\mathcal{D}|}} (d) \mathrm{d} P(\xi_1,\ldots ,\xi_{|\mathcal{D}|}).
\end{eqnarray*}
For any function $f(.)$, the marginal expectation of the sample mean, averaging over both sampling and distributional uncertainty, is
\begin{equation}\label{eq:mean}
    \mathbb{E}_\text{marginal} \left[\frac{1}{n} \sum_{i=1}^n f(D_i) \right] = \mathbb{E}_P[f(D)]
\end{equation}
and the marginal variance of the sample mean is
\begin{equation}\label{eq:variance-formula}
   \text{Var}_\text{marginal} \left(\frac{1}{n} \sum_{i=1}^n f(D_i) \right) = \frac{\delta^2}{n} \text{Var}_P(f(D)),
\end{equation}
where the scaling factor $\delta^2$ satisfies
\begin{equation}\label{eq:def-delta2}
 \frac{\delta^2}{n}=  \underbrace{\frac{1}{n}}_{\substack{\text{due to sampling}}} +  \underbrace{ \text{Var}(\xi_1) \frac{n-1}{n} \frac{|\mathcal{D}|^2}{|\mathcal{D}|-1} }_{\substack{\text{due to distributional perturbation}}}.
\end{equation}
Note that we write for simplicity the sub-index "$\text{marginal}$" instead of $P_{\text{marginal}}$. Under regularity assumptions \cite[Sec.2]{jeong2022calibrated} one can show 
\begin{equation}\label{eq:asympnorm}
    \frac{1}{n} \sum_{i=1}^n f(D_i) - \mathbb{E}_P[f(D)] \approx \mathcal{N} \left( 0, \frac{\delta^2}{n} \text{Var}_P(f(D)) \right).
\end{equation}
Let us give some intuition on how to interpret different values of $\delta$. By equation~\eqref{eq:def-delta2}, $\delta^2 \ge 1$. If $\delta=1$, then $\text{Var}(\xi_i) = 0$ and thus there is no distributional perturbation, that is the data is drawn i.i.d.\ from $P^\xi = P$. As $\delta$ increases, the $f(D_i)$ become increasingly correlated marginally. 
The variance $\text{Var}(\xi_1)$ is maximized for $\xi_i \in \{ 0,1 \}$ for all $i$. In this case, $\text{Var}(\xi_1) = \frac{1}{|\mathcal{D}|} - \frac{1}{|\mathcal{D}|^2}$. Using equation~\eqref{eq:def-delta2} in this most extreme case, we get $\delta^2 = n$. Overall, we have the bound
\begin{equation*}
    1 \le \delta^2 \le n.
\end{equation*}

\subsubsection{A numerical example.}

\begin{figure}
\begin{center}
    \includegraphics[scale=.45]{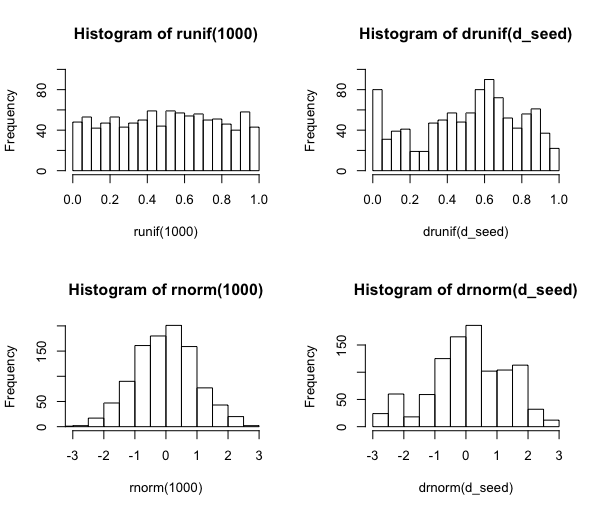}
\end{center}
\caption{Random number generation from the distributional perturbation model. On the upper left, the observations are drawn i.i.d.\ from the uniform distribution. On the upper right, the observations are drawn from a perturbed uniform distribution. On the bottom left, the observations are drawn from the standard Gaussian distribution. On the lower right, the observations are drawn from the perturbed Gaussian distribution. In each case, the sample size is $n=1000$. For the distributional perturbations we use $\delta = 5$.}\label{fig:simulate_data}
\end{figure}

The R package calinf (\url{github.com/rothenhaeusler/calinf}) contains functions to generate data from perturbed distributions. Sampling from the distributional perturbation model is slightly more involved than drawing i.i.d.\ random variables, since we have to choose the strength of the perturbation.  %
We set the state of the distributional perturbation by setting a distributional seed via \texttt{distributional\_seed}. This step is not optional.
On a high level, the distributional seed indicates to the random number generator which observations are drawn from the same perturbed distribution, allowing the random number generator to introduce spurious associations between the variables.  %
Once the distributional seed is set, one can use this to generate perturbed data as follows. 
\begin{verbatim}
d_seed <- distributional_seed(n=1000,
                            delta=5)
x <- drnorm(d_seed)
y <- drnorm(d_seed)    
\end{verbatim}
The displayed code generates 1000 observations from $P^\xi$, where $P^\xi$ is a perturbed two-dimensional standard Gaussian distribution. %
The perturbed data is generated such that equation~\eqref{eq:variance-formula} holds approximately for any square-integrable $f(D)$. For continuous random variables, one cannot directly use the strategy described in Section~\ref{sec.distrpert}. Details on how to sample from perturbed continuous distributions can be found in \cite[Sec.2]{jeong2022calibrated}.
The function \texttt{drnorm} generates i.i.d.\ data from a perturbed Gaussian. 
Analogously, we provide functions to sample from a perturbed binomial distribution (\texttt{drbinom}), perturbed uniform distribution (\texttt{drunif}), etc. Here, the "dr" in drunif stands for "distributional randomness". An example is shown in Figure~\ref{fig:simulate_data}.

\subsection{Estimation under the distributional perturbation model}

In this section we describe how to do estimation and inference in the distributional perturbation model from Section \ref{sec.distrpert}.

In a nutshell, as the expectation of sample means is unchanged, estimation under the distributional perturbation model can  proceed "as usual": irrespective of the value of $\delta$, one can construct point estimators such as moment-based or
maximum-likelihood estimators as if the data were drawn i.i.d.\ from the target distribution $P$. %

As an example, let us 
focus on the OLS parameter $\theta(P) = \arg \min \mathbb E_{P}[(Y - X \theta )^2]$. 
If $\theta(P)$ is unique,
it can be rewritten as
\begin{equation*}
    \theta(P) = \mathbb{E}_P[ X^\intercal X]^{-1} \mathbb{E}_P[X^\intercal Y].
\end{equation*}
Assume that $D_i = (X_i,Y_i)$, $i=1,\ldots,n$ are drawn i.i.d.\ from $P^\xi$. Due to equation \eqref{eq:mean}, marginally across the sampling and distributional perturbation, we have
\begin{align*}
  \mathbb{E}_\text{marginal}  \left[ \frac{1}{n} \sum_{i=1}^n X_i^\intercal X_i \right] &= \mathbb{E}_P[X^\intercal X] , \text{ and } \\
      \mathbb{E}_\text{marginal} \left[\frac{1}{n} \sum_{i=1}^n X_i^\intercal Y_i \right] &= \mathbb{E}_P[X^\intercal Y].
\end{align*}
This motivates the estimator
 \begin{equation*}
     \hat \theta =  \left( \frac{1}{n} \sum_{i=1}^n X_i^\intercal X_i \right)^{-1}  \left( \frac{1}{n} \sum_{i=1}^n X_i^\intercal Y_i \right),
 \end{equation*}
 which is the usual OLS estimator that we would use if the data $(X_i,Y_i)_i$ were drawn i.i.d.\ from $P$. %
 Similar ideas can be applied to maximum likelihood estimation and the method of moments to show that estimation can proceed "as usual"
 \cite{jeong2022calibrated}. 
 
 On the other hand, as we will discuss 
 below, the variance of the resulting estimator depends on the (usually unknown) $\delta$. Under the usual and additional minor assumptions, such estimators are asymptotically linear and Gaussian under the distributional perturbation model \cite{jeong2022calibrated}.  For the example with OLS regression, a Taylor expansion shows that
\begin{eqnarray}\label{eq:linear-expansion}
\hat \theta - \theta(P) = 
    & &\frac{1}{n} \sum_{i=1}^n \underbrace{ \mathbb{E}_P [X^\intercal X]^{-1} X_i^\intercal (Y_i - X_i \theta(P))}_{\substack{ \phi_{P}(D_i) }}\nonumber\\
    & &\hspace*{39mm} + o_{P_\text{marginal}}(\frac{\delta}{\sqrt{n}}).
\end{eqnarray}
Thus, up to lower order terms, 
the difference between the estimator and the target parameter (computed on the unperturbed distribution) is a mean of a function of the data. This is similar to classical expansions in terms of the influence function \citep{van2000asymptotic}. The main difference is that the expansion is done in a non-i.i.d.\ setup that accounts for both sampling uncertainty and distributional uncertainty. Since the estimator is asymptotically linear, we can apply equation~\eqref{eq:asympnorm} to equation~\eqref{eq:linear-expansion}. Thus, under regularity assumptions \cite{jeong2022calibrated}, one obtains 
\begin{equation*}
    \hat \theta - \theta(P) \approx \mathcal{N} \left(0, \frac{\delta^2}{n} \text{Var}_P(\phi_{P}(D)) \right),
\end{equation*}
where the distributional approximation is meant to hold w.r.t. $P_\text{marginal}$ on the left-hand side. %
In the following, we will see that standard approaches will fail at estimating the correct variance of $\hat \theta$.

\subsubsection{Classical statistical inference may drastically underestimate uncertainty.}

In the model above, one might be tempted to estimate the variance of statistical quantities as usual. Let's consider the example of the sample mean $\overline D = \frac{1}{n} \sum_{i=1}^n D_i$. In the following, we will see that it is straightforward to estimate $\text{Var}_P(D)$, but that estimation of $\text{Var}_\text{marginal}(\overline D)$ is more challenging.

\paragraph*{Estimation of $\text{Var}_P(D)$.} If the data were drawn i.i.d.\ from the target distribution $P$, one would use the variance estimator $ \hat \sigma^2$, where $\hat \sigma^2$ is the empirical variance
\begin{align*}
 \hat \sigma^2 &=  \frac{1}{n} \sum_{i=1}^n (D_i - \frac{1}{n} \sum_{j=1}^n D_j)^2 \\
 &=  \frac{1}{n} \sum_{i=1}^n D_i^2 - ( \frac{1}{n} \sum_{i=1}^n D_i )^2 
\end{align*}
Let us now investigate the variance estimator $\hat \sigma^2$. We will see that $\mathbb{E}_\text{marginal}[\hat \sigma^2]$ is close to $\text{Var}_P(D)$. 
By equation~\eqref{eq:mean}, the marginal expectation of $\frac{1}{n} \sum_i D_i^2$ is $\mathbb{E}_P[D^2]$. Similarly, the marginal expectation of $\overline D$ is $\mathbb{E}_P[D]$. Using equation~\eqref{eq:variance-formula}, the variance of $\overline D$ is $\delta^2 \text{Var}_P(D)/n$. Thus, the empirical variance estimate will have expected value
\begin{align*}
&\mathbb{E}_\text{marginal}[\hat \sigma^2] \\
&=    \mathbb{E}_P[D^2] - ( \mathbb{E}_P[D]^2 + \frac{\delta^2}{n} \text{Var}_P(D)) \\
&= \left(1- \frac{\delta^2}{n} \right)\text{Var}_P(D) 
\end{align*}
Thus, if $\delta^2$ is small or $n$ is large, then $\mathbb{E}_\text{marginal}[\hat \sigma^2]$ is close to $\text{Var}_P(D)$; that is the difference is negligible. 

This effect can also be easily observed empirically, here illustrated with some commands 
in \texttt{R}. 
In the following, we draw $n=1000$ observations in a distributional perturbation model with $\delta=2$. The estimated variance is relatively close to the variance $\text{Var}_P(D) =1 $, where $P = \mathcal{N}(0,1)$.
\begin{verbatim}
> d_seed <- distributional_seed(
                        n     =   1000,
                        delta =   2
                        )
> D <- drnorm(d_seed)
> var(D)
[1] 0.9414752
\end{verbatim}
This is good news. However, there are also some bad news. To construct confidence intervals for $\mathbb{E}_P[D]$, we need an estimator of the variance of $\overline D = \frac{1}{n} \sum_{i=1}^n D_i$.

\paragraph*{Naive estimation of $\text{Var}_\text{marginal}(\overline D)$.}
 As we will see in the following, the naive estimator $\hat \sigma^2_\text{naive} = \frac{1}{n} \hat \sigma^2$ systematically underestimates the variance of $\overline D$, potentially drastically so. 
Intuitively, this is the case because the data points $D_i$ are positively correlated under $P_\text{marginal}$, with unknown correlation. Let us compute the expectation:
\begin{align*}
    \mathbb{E}_\text{marginal}[ \hat \sigma_{\text{naive}}^2] &= \frac{1}{n} \left(1- \frac{\delta^2}{n} \right)\text{Var}_P(D)  &  \\
    &< \frac{\delta^2}{n} \text{Var}_P(D) \qquad &\text{equation \eqref{eq:def-delta2}}\\
    &=  \text{Var}_\text{marginal}(\overline D) \qquad &\text{equation \eqref{eq:variance-formula}}
\end{align*}
Thus, $\hat \sigma_{\text{naive}}^2$ systematically underestimates  $\text{Var}_\text{marginal}(\overline D)$. As discussed before, in the most extreme case $\delta^2 =n$ which would make the left-hand-side equal to zero. We can also see this effect empirically, illustrated in \texttt{R} below. The naive estimator, computed on the previous example, is
\begin{verbatim}
> var(D)/n
[1] 0.0009414752
\end{verbatim}
On the other hand, the actual variance of $\overline D$, marginally across both the distributional perturbation and the sampling process is
\begin{verbatim}
> simulate_mean <- function(){
> d_seed <- distributional_seed(
                        n     =   1000,
                        delta =   2
                        )
> D <- drnorm(d_seed)
> return(mean(D))
> }
> var(replicate(n=10000,simulate_mean()))
[1] 0.003949516
\end{verbatim}
Thus, the naive estimator $\hat \sigma_{\text{naive}}^2$ underestimates the variance roughly by a factor of $4$ 
which is to be expected since $\delta^2 = 4$.

Summarizing the discussion in this section, estimation of $\sigma^2 = \text{Var}_P(D)$ can be done as usual, while estimation of $\text{Var}_\text{marginal}(\overline D)$ is more difficult. To be more specific, one can use the empirical variance of $(D_i)_{i=1,\ldots,n}$ to estimate $\sigma^2 = \text{Var}_P(D)$. In the more general case of asymptotically linear estimators, one can estimate the variance $Var_P(\phi_P(D))$ by computing the empirical variance of $(\hat \phi(D_i))_{i=1,\ldots,n}$, where $\hat \phi$ is a plug-in estimate of the influence function \cite{jeong2022calibrated}. Let us now turn to estimation of $\text{Var}_\text{marginal}(\overline D)$. Since 
 \begin{equation*}
     \text{Var}_\text{marginal}(\overline D) = \frac{\delta^2}{n} \text{Var}_P(D),
 \end{equation*}
 the main challenge is to estimate $\delta$. In the following two sections we will discuss two approaches to estimate $\delta$.  %

\subsubsection{Calibration of uncertainty using triangulation.}\label{sec:triangulation}

In this section we discuss how a commonly recommended research strategy, called "method triangulation" can be used to estimate distributional uncertainty . 

If several estimators of an effect are available, one can use variation of the estimators as a measure of robustness. %
In the statistics literature, this type of stability analysis has been advocated
by \citet{yu2020} as part of the predictability, computability, and stability (PCS) framework. More generally speaking, investigating stability across methods is often referred to as method triangulation \citep{denzen1978sociological,patton1999enhancing,munafo2018repeating}.  Triangulation is conceptually different from replicability across settings. For example, if the same study is conducted multiple times at different locations, these studies may share similar biases and thus may be consistently incorrect. On the other hand, if different methodologies yield similar conclusions, then the result is less likely to be an artifact. These intuitive arguments can be made precise in the distributional uncertainty framework. %

Assume we have access to several estimators $\hat \theta_1, \ldots, \hat \theta_K$ 
for the same
parameter of interest $\theta(P)$. Examples from causal inference include settings where we have
\begin{itemize}
\item multiple instruments,
\item multiple adjustment sets, or
\item treatment effect homogeneity.
\end{itemize}
For example in presence of %
treatment effect homogeneity, we can estimate average treatment effects on various subpopulations. If there were no distributional uncertainty across the subpopulations, these estimators should agree, at least asymptotically. On the other hand, if there is a lot of distributional uncertainty, these estimators will be very far apart from each other. Thus, 
we can use the observed variation between estimators as an indication of how much distributional uncertainty is present for the problem at hand. In the following, we will make this more precise.

 We assume that the estimators are asymptotically linear, that is,
\begin{equation*}
    \hat \theta_k - \theta_k(P) = \frac{1}{n} \sum_{i=1}^n \phi_k(D_i) + o_{P_{\text{marginal}}}(\frac{\delta}{\sqrt{n}}),
\end{equation*}
for some mean-zero functions $\phi_k$, that is $\mathbb{E}_P[\phi_k(D)]=0$.
For the example of ordinary least squares estimation, see also equation~\eqref{eq:linear-expansion}.
This is also justified for maximum likelihood estimators and empirical risk minimization in low-dimensional settings, see \cite{jeong2022calibrated}.
For simplicity, in the following we will assume that $\theta_k(P) = \theta_{\ell}(P)$ for all $k,\ell$. 
This corresponds to the assumption that if no uncertainty were present (infinite data from the target distribution), all estimators would return the same target quantity. One might have reasons to doubt this assumption. If this assumption holds, the inferential procedure described below will have exact coverage asymptotically. If it is violated, one will generally have overcoverage \citep{jeong2022calibrated}.

Now let us proceed with the estimation of $\delta$. 
The variation between the different estimation strategies is a measure of the trustworthiness of the result. %
Considering the squared difference of the estimators yields
\begin{align}\label{eq:chisq}
\begin{split}
    &n(\hat \theta_k - \hat \theta_{\ell})^2  \\
      & = n( \frac{1}{n} \sum_{i=1}^n \phi_k(D_i) - \phi_{\ell}(D_i))^2 + o_{P_{\text{marginal}}}(\delta^2) \\
     &\approx \delta^2 \text{Var}_P(\phi_k(D) - \phi_{\ell}(D)) \chi^2_1 
\end{split}
\end{align}
Here, we used equation~\eqref{eq:asympnorm}.
Thus, we can form an %
estimate of $\delta$ by setting
\begin{equation}\label{eq:estimate}
  \hat \delta^2 = \frac{1}{K(K-1)} \sum_{k \neq \ell} \frac{n(\hat \theta_k - \hat \theta_{\ell})^2}{\widehat{\text{Var}}_P(\phi_k(D) - \phi_{\ell}(D))}.
\end{equation}

One can then use this estimate in conjunction with equation~\eqref{eq:asympnorm} to form 95\%-confidence intervals:
\begin{equation}\label{eq:calibrated-ci}
    \hat \theta  \pm  1.96 \frac{\hat \delta \hat \sigma}{\sqrt{n}}
\end{equation}
Here, $\hat \sigma^2$ is the usual variance estimate one would use if the data were drawn i.i.d.\ from the target distribution. More specifically, one can estimate the influence function $\phi$ of $\hat \theta$ and use plug-in estimate of the variance: 
\begin{equation}\label{eq:variance-est}
\hat \sigma^2 = \frac{1}{n-1} \sum_{i=1}^n (\hat \phi(D_i) - \frac{1}{n} \sum_{j=1}^n \hat \phi(D_i))^2.
\end{equation}

Under regularity assumptions and for large $K$, this interval is valid in an asymptotic sense \citep{jeong2022calibrated}:
\begin{equation}\label{eq:calibrated-guarantee}
    \mathbb{P}_\text{marginal} \left[ |\theta(P) -\hat \theta| \le   z_{1-\alpha/2} \frac{\hat \delta \hat \sigma}{\sqrt{n}} \right] \rightarrow 1-\alpha,
\end{equation}
where $z_{1-\alpha/2}$ is the $1-\alpha/2$-quantile of a standard Gaussian random variable. If $K$ is small, then  equation~\eqref{eq:chisq} suggests replacing $z_{1-\alpha/2}$ with quantiles of a $t$-distribution with appropriate degrees of freedom \citep{jeong2022calibrated}. The main takeaway here is that we give coverage guarantees for the unperturbed parameter $\theta(P)$, as opposed to the perturbed parameter $\theta(P^\xi)$. 
Furthermore, these guarantees hold marginally, that means across multiple draws of both the distributional and sampling uncertainty. 

An important aspect that we have glossed over until now is that for this procedure to work the estimators $\hat \theta_k$ have to be sufficiently different. As an extreme example, one cannot use $\hat \theta_1 = \hat \theta_2 = \ldots = \hat \theta_K$. With "sufficiently different" we mean that the influence functions of the estimators have to be different. This is reflected in equation~\eqref{eq:estimate}. If the influence functions are very similar, the denominator in \eqref{eq:estimate} goes to zero and the procedure becomes increasingly unstable. More details can be found in \cite{jeong2022calibrated}.

One important takeaway from this methodology is that it is not the absolute stability (empirical variation of the estimators) that matters, but relative stability. In equation~\eqref{eq:estimate}, we divide the variation $(\hat \theta_k - \hat \theta_{\ell})^2$ by the expected variation under i.i.d.\ sampling $\frac{1}{n} \widehat{\text{Var}}_P( \phi_k(D) -  \phi_{\ell}(D))$. If the actual variation is larger than the expected variation under i.i.d.\ sampling, we have some indication that there is distributional uncertainty. %

\subsubsection{Calibration of uncertainty using knowledge about the superpopulation.}\label{sec:moment-eq}

Knowledge about the superpopulation can be leveraged to estimate $\delta$. As an example, the data scientist might know the average age or average income of the target population. Such knowledge can be expressed as moment equations. If the empirical average age is far from the target population average age, then this is an indication that either distribution or sampling uncertainty is high. Thus, we can use such knowledge to construct an estimator of $\delta$. Let us now formalize this idea. As an example, assume that we know
\begin{equation*}
    \mu = \mathbb{E}_{P}[X],
\end{equation*}
where $\mu \in \mathbb{R}^K$.
In this case, using equation~\eqref{eq:variance-formula}, for any fixed $k$ we can construct an unbiased estimate of $\delta^2 \text{Var}_{P}(X)$: %
\begin{equation*}
     n(\overline X_{\bullet k} - \mu_k )^2.
\end{equation*}
Similarly, for each $k$ we can construct an estimator of $\delta^2$ by setting
\begin{equation*}
    \hat \delta_k^2 = \frac{ n (\overline X_{\bullet k} - \mu_k )^2}{\frac{1}{n-1} \sum_{i=1}^n (X_{ik} - \overline X_{\bullet k})^2}
\end{equation*}
Note that this is the squared $t$-test statistic. Even for $n \rightarrow \infty$, the variance of $\hat \delta_k^2$ does not go to zero. Under the non-i.i.d.\ sampling model, using equation~\eqref{eq:asympnorm}, $\hat \delta_k^2/ \delta^2$ follows a $\chi^2_1$-distribution asymptotically. 
Precision can be gained by averaging:
\begin{equation*}
    \hat \delta^2 = \frac{1}{K} \sum_{k=1}^K \hat \delta^2_k.
\end{equation*}
This estimate of $\delta$ can then be used to construct confidence intervals as described in
equation~\eqref{eq:calibrated-ci}. Under appropriate regularity assumptions, $\hat \delta \rightarrow \delta$. Thus, this approach will yield asymptotic coverage guarantees as in equation~\eqref{eq:calibrated-guarantee}, see \cite{jeong2022calibrated}.

\subsection{Calibrated inference in R}\label{sec:calibR}

\begin{figure}
    \centering
    \includegraphics[scale=.6]{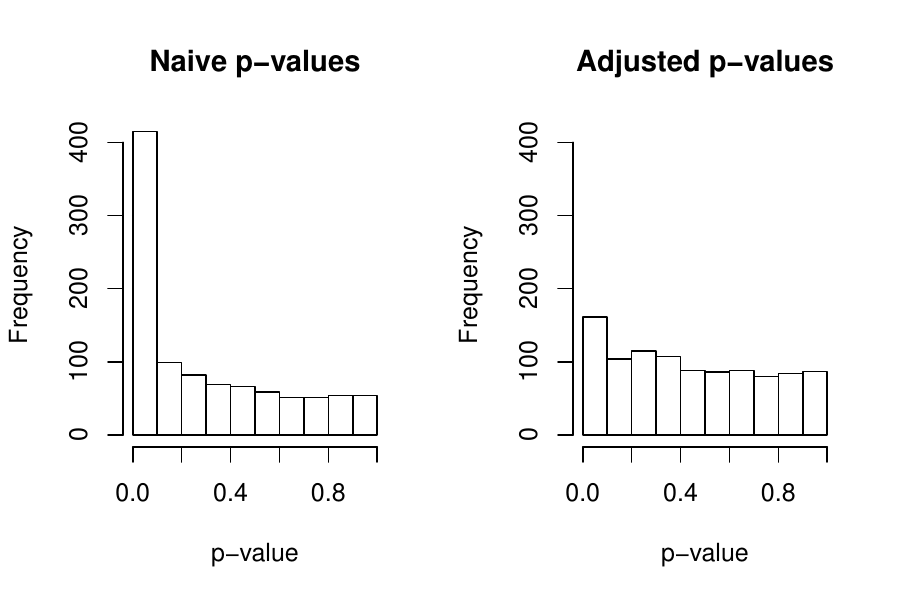}
    \caption{Example from Section~\ref{sec:calibR}. On the left-hand side, we show the histogram of $N= 1000$ naive $p$-values as reported by \texttt{lm()}. On the right-hand side, the $p$-values are computed via \texttt{calm()}, i.e.\ the $p$-values are calibrated. The null hypothesis $\theta(P) = 0$ is true. As expected, the naive $p$-values are not valid for this hypothesis. In fact, more than 40\% of the naive $p$-values are smaller than $.1$. While not perfect, the distribution of the adjusted $p$-values is much closer to a uniform distribution.}
    \label{fig:histogram-pval}
\end{figure}

In the following, we describe some functions available in the R-package available on GitHub (\url{github.com/rothenhaeusler/calinf}) that allow to quantify both sampling and distributional uncertainty. At the center is the approach described in Section~\ref{sec:triangulation}. As an example, let us consider %
the problem of estimating the causal effect of some binary treatment $Tr \in \{0,1 \}$ %
on some outcome $Y$ via linear regression, in the presence of some covariates $X_1,\ldots,X_5$. The practitioner might have several reasonable choices for confounder adjustment. 
Examples of variables that can (but are not necessarily included) in regression adjustment are exogeneous variables that affect the outcome, but not the treatment. Similarly, instrumental variables affect the treatment but are assumed to have no direct effect on the outcome. For valid treatment effect estimation, such adjustment variables can be (but do not have to be) included in a regression. 
These choices can be specified in a list of formulas:
\begin{verbatim}
formulas <- list(Y ~ Tr + X1 + X2, 
                 Y ~ Tr + X1 + X2 + X3,
                 Y ~ Tr + X1 + X3 + X4,
                 Y ~ Tr + X1 + X2 + X5
                 )
\end{verbatim}
In a second step, one can then run a calibrated linear regression:
\begin{verbatim}
    calm(formulas, data = data, 
            target = "Tr")
\end{verbatim}

Let us consider a concrete numerical example. We are interested in estimating the causal effect of a binary treatment variable $Tr$ on $Y$ in a structural causal model \citep{pearl2009causality,peters2017elements}.  Let $P$ be 
the distribution of $(Tr, I_1, X_1, X_2, J_1, J_2, Y)$ which
is generated as follows:
\begin{align*}
    X_1 & = \epsilon_1 \\
    X_2 & = X_1 + \epsilon_2 \\
    I_1 &= \epsilon_3 \\
    J_1 &= \epsilon_4 \\
    J_2 &= J_1 + \epsilon_5 \\
    Tr &= X_1 + X_2 + I_1 +\epsilon_6 \\ 
    Y &= Tr + X_1 - X_2 + J_1 + J_2 + \epsilon_7
\end{align*}
Here, $\epsilon \sim \mathcal{N}(0, \text{Id}_7)$. In words, $I_1$ is an instrument and $(J_1,J_2)$ are variables that affect $Y$ but not the treatment. We are interested in the direct causal effect of $Tr$ on $Y$, which in this setting can be written as $\theta(P) = \arg \min_\theta \min_\beta \mathbb{E}_P[(Y -  Tr \cdot \theta - 
Z \beta)^2]$ for some appropriate set of adjustment variables $Z$. In this setting, there are multiple valid estimation strategies for $\theta(P)$. More precisely, all of the following formulas are valid in the sense that if one had infinite data from $P$, regression adjustment via these formulas would yield a consistent estimator of $\theta(P)$:
\begin{verbatim}
formulas <- list(
    Y ~ Tr + X1 + X2, 
    Y ~ Tr + X1 + X2 + I_1,
    Y ~ Tr + X1 + X2 + J_1,
    Y ~ Tr + X1 + X2 + J_2,
    Y ~ Tr + X1 + X2 + J_2 + I_1,
    Y ~ Tr + X1 + X2 + J_1 + I_1,
    Y ~ Tr + X1 + X2 + J_1 + J_2,
    Y ~ Tr + X1 + X2 + J_1 + J_2 + I_1
                 )
\end{verbatim}
We sample $n = 100$ observations from the random perturbation model with $\delta = 2$. The value $\delta$ is not known to the data scientist and thus has to be estimated. Running calibrated linear regression yields the following output:
\begin{verbatim}
> calm(formulas,df,target="Tr")

Quantification of both distributional 
and sampling uncertainty

   Estimate Std. Error Pr(>|z|)
Tr   1.0157     0.0676        0

hat delta = 2.376035
\end{verbatim}
As we can see, the estimated scaling factor $\hat \delta$ is somewhat close to $\delta = 2$. Estimation of $\delta$ is somewhat unstable across draws from the perturbation model. This is due to the fact that estimation of $\hat \delta$ has non-negligible variance even for $n \rightarrow \infty$ (see  equation~\eqref{eq:chisq}). Precision of $\hat \delta$ can be improved by adding additional estimators or moment constraints. From a statistical perspective, it is pertinent to investigate the validity of $p$-values across both sampling uncertainty and distributional uncertainty. To investigate the validity of $p$-values, we set the direct causal effect in the structural equation model to zero, that is, we set
\begin{equation*}
        Y = X_1 - X_2 + J_1 + J_2 + \epsilon_7.
\end{equation*}
We then compute naive $p$-values as reported by
\begin{verbatim}
lm(Y ~ Tr + X1 + X2)
\end{verbatim}
and also compute calibrated $p$-values via
\begin{verbatim}
calm(formulas,df,target="Tr").
\end{verbatim}
We repeat the two-stage sampling and estimation procedure $N = 1000$ times. The histograms of $p$-values are depicted in Figure~\ref{fig:histogram-pval}. 

The naive $p$-values are not valid for the hypothesis $\theta(P) = 0$, due to the distributional uncertainty. If there were no distributional uncertainty, the naive $p$-values would be valid. Intuitively speaking, the naive $p$-values are based on a variance formula that drastically underestimates uncertainty for the parameter
$\theta(P)$. Thus, these $p$-values are anti-conservative.  The $p$-values as reported by calm, while not perfect, follow roughly a uniform distribution.

The \texttt{R}-package \texttt{calinf} provides functions also for calibrating inference in generalized linear models. If the outcome $Y$ is binary, one can run calibrated logistic regression:
\begin{verbatim}
caglm(formulas, family = "binomial", 
        data=data, target="Tr")
\end{verbatim}
The function \texttt{caglm} is a wrapper for \texttt{glm}. Thus, one can run any generalized linear model by specifying an appropriate family in \texttt{caglm}. 

Looking further, the proposed procedure in Section~\ref{sec:triangulation} is not limited to calibrate uncertainty only for generalized linear models. In principle, the proposed approach can be used for any asymptotically linear estimators. In the future, we aim to provide additional functionality that extend beyond these simple use cases.

\subsection{Uniqueness of the distributional uncertainty model.}

The discussion in the previous sections raises the question whether there are other non-adversarial perturbation models that would have led to different asymptotics. In this section, we give a negative answer to this question, within the assumed framework of a randomly perturbed distribution $P^{\xi}$. 

In the following, for each realization of $\xi$ let $P^\xi$ be a probability measure on $\mathcal{D}$. To be more specific, we assume that $P^\bullet$ is a random probability measure. As an example, $P^\xi$ might be constructed via random re-weighting with potentially non-exchangeable $\xi_i$'s. In the following we will assume that $P^\xi$ is  "unbiased", i.e.\ that for every measurable set $A \subseteq \mathcal{D}$ we have $E_\xi[P^\xi[D \in A]] = P[D \in A]$.

When considering distributional perturbation models, arguably there are two assumptions that may seem natural. First, one would like to have that events with probabilities close to zero are only perturbed very little (otherwise, $P^\xi$ would be very different from $P$). To be more precise, we require that for every sequence of measurable sets $A_1, A_2, \ldots \subseteq \mathcal{D} $ with
\begin{equation*}
    P(D \in A_j) \rightarrow 0\ (j \to \infty)
\end{equation*}
we have
\begin{equation}\label{eq:assumption1}
    \text{Var}_{\xi}(P^\xi( D \in A_j)) \rightarrow 0\ (j \to \infty).
\end{equation}

In addition, we would like to exclude adversarial perturbations that only change a distribution in a very specific way. Mathematically, we model this by an isotropic perturbation, that means that events that have equal probability, are perturbed similarly. This can be seen as a symmetry assumption. To be specific, if $P(D \in A) = P(D \in B)$ then we assume that
\begin{equation}\label{eq:assumption2}
    \text{Var}_{\xi}(P^\xi(D \in A)) = \text{Var}_{\xi}(P^\xi(D \in B)).
\end{equation}

\begin{theorem}[\cite{jeong2022calibrated},Th.2]\label{th1}
Assume that \eqref{eq:assumption1} and \eqref{eq:assumption2} holds. Furthermore, assume that there exists a measurable function $u(D)$ such that $u(D) \sim  \text{Unif}([0,1])$, for $D \sim P$.
Then, there exists $\delta_\text{dist} \ge 0$ such that for any square-integrable function $f(D) \in L^2(P)$,
\begin{equation*}
  \mathrm{Var}_\xi(\mathbb{E}_\xi[f(D)]) = \delta_\text{dist}^2 \mathrm{Var}_{P}(f(D)).
\end{equation*}
\end{theorem}

The implication of Theorem~\ref{th1} is as follows. 
Assume that conditionally on $\xi$, the data $(D_i)_{i=1,\ldots,n}$ is drawn i.i.d.\ from the perturbed distribution $P^\xi$. 
Then, for all square-integrable functions $f(D) \in L^2(P)$ we have
\begin{align*}
  &\text{Var}_\text{marginal} ( \frac{1}{n} \sum_{i=1}^n f(D_i) ) \\
  =\, \, &  (\frac{1}{n}  + \delta_\text{dist}^2 - \frac{\delta_\text{dist}^2}{n}) \text{Var}_{P}(f(D)).
\end{align*}
Ignoring the lower-order term $\frac{\delta_\text{dist}^2}{n}$, we can combine uncertainty due to sampling and uncertainty due to the distributional perturbation by setting
\begin{equation*}
    \delta^2 = 1 + n \delta_\text{dist}^2.
\end{equation*}
Then, for all square-integrable functions $f(D) \in L^2(P)$ marginally across both sampling and distributional uncertainty we have
\begin{equation*}
    \text{Var}_\text{marginal} ( \frac{1}{n} \sum_{i=1}^n f(D_i) ) \approx  \frac{\delta^2}{n} \text{Var}_{P}(f(D)).
\end{equation*}
This corresponds to the perturbation model introduced in equation~\eqref{eq:variance-formula}.

\section{Application}\label{sec:real-world}

We apply calibrated inference to a Get-Out-The-Vote field experiment, which investigates whether voter turnout can be increased by social pressure \citep{gerber2008social}. We study two 
groups: the "control" group and the "neighbors" group, and we refer to the latter also as the treatment group. %
The "neighbors" group received a mail with the statement "DO  YOUR  CIVIC  DUTY—VOTE!". The letter lists the voting record of neighbors and threatens to publicize who does and does not vote. The outcome is voter turnout in the August 2006 primary election in Michigan.

The treatment is applied on the household level. On average, there are approximately $2$ units per household. Since units within households are correlated, the data should be analyzed using clustered standard errors. This data generation process can also be seen as a random perturbation model, where units only appear in the dataset if all other units in the household also appear in the dataset. That is, the treatment group is observed from a distribution which is different from the idealized one with i.i.d. sampling for which we want to infer the treatment effect. We want to emulate a scenario where the data is not drawn i.i.d.\ from the target distribution, with unknown correlations between units. Thus, we drop the household indicator, and hope to recover valid inferential statements by calibrating the $p$-values. 

Since the ground truth is unknown, we re-randomize the treatment variable to simulate a setting where the treatment effect is zero. The covariates and outcomes are left unchanged.

In this setup, one expects the correlation within household units to inflate the variance compared to i.i.d.\ sampling. In our scenario for illustration, as mentioned above, the household indicator is considered unknown. Thus, we have to infer the variance inflation factor $\delta$ from data alone. To estimate $\delta$, we use super-population constraints as described in Section~\ref{sec:moment-eq}. To form these constraints, we use that for each individual we have records whether they voted in the primary elections in 2000, 2002, and  2004 or the general election in 2000 and 2002. 
For each of these covariates, as super-population constraints we assume that 
the covariance between treatment and covariates is zero. Intuitively, if the empirical covariance between treatment and covariates is significantly different from zero under an
i.i.d.\ sampling, there is evidence of positive associations between units. 

\begin{figure}
    \centering
    \includegraphics[scale=.6]{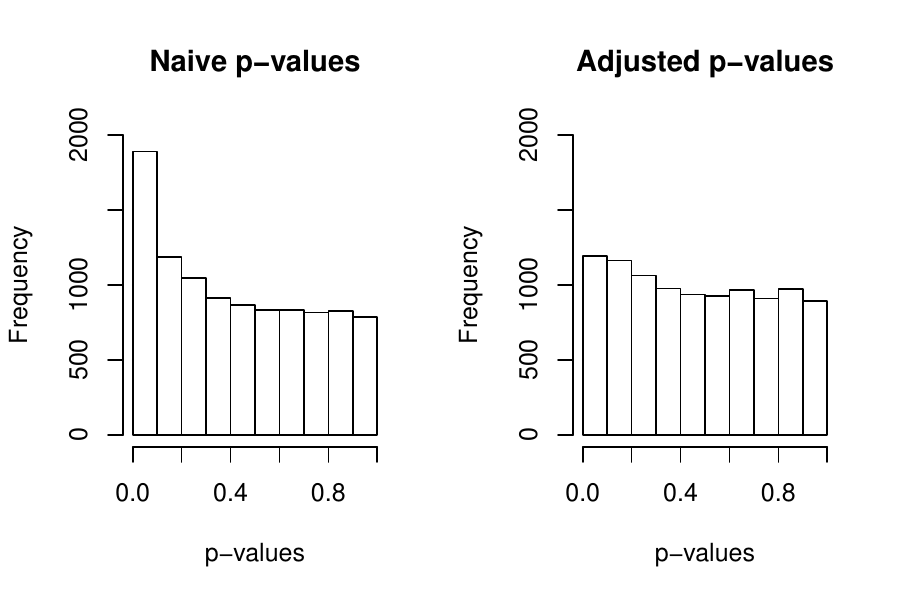}
    \caption{Calibrated inference for the Get-Out-The-Vote field experiment \citep{gerber2008social}. On the left-hand side, we show the histogram of $N= 1000$ naive $p$-values computed via difference-in-means. On the right-hand side, the $p$-values are calibrated using super-population constraints. The treatment has been re-randomized to guarantee that the null hypothesis $\theta(P) = 0$ is true. Thus, the $p$-values should follow a uniform distribution. As some of the units are positively correlated, the naive $p$-values are not valid.  Around 11\% of the naive $p$-values are below $.05$ while only 6.5\% of the adjusted $p$-values are below $.05$. The empirical distribution of adjusted $p$-values is much closer to a uniform distribution.} 
    \label{fig:GOTV}
\end{figure}

There are $n = 119,999$ households in the dataset that were subject to the treatment or control group.
We randomly select $m = 1,200 \approx n/100$ households and compute calibrated $p$-values as well as naive $p$-values via difference-in-means, assuming that the household identifier is unknown. This process was repeated $10,000$ times. The resulting $p$-values are depicted in Figure~\ref{fig:GOTV}. The calibrated $p$-values follow 
much closer 
a uniform distribution (which is correct) than the naive $p$-values. Around 11\% of the naive $p$-values are below $.05$ while only 6.5\% of the calibrated $p$-values are below $.05$. This indicates that the calibration procedure succeeded at capturing the excess variation due to unobserved clustering.

\section{Discussion and outlook}

We summarize the main points of our exposition 
and outline how the propagated ideas can potentially be extended to improve replicability and generalizability.

In many practical problems, the data is not drawn i.i.d.\ from the target population. For example, unobserved sampling bias, confounding, batch effects, or unknown associations can inflate the deviation of the estimator from its target compared to i.i.d.\ sampling. For reliable statistical inference, it is of paramount importance 
to account for these additional types of uncertainty. Failure to do so is a major source of lack of replicability of scientific findings in many fields. 

We present two approaches
to deal with such distribution shifts. In Section~\ref{sec:externalstability} we consider a directional notion of distributional stability. In the existing literature, distributional errors are often handled via worst-case bounds. Such bounds can be very conservative and  may lead to rather limited information gain as some type of shifts might be more realistic than others. The directional notion of stability allows to probe different perturbations to investigate what type of distribution shift the estimand is most sensitive to. This then leads to a less conservative notion of sensitivity and helps to judge which type of distribution shifts one should be most worried about.

In Section~\ref{sec:calinf} we go beyond worst-case stability. All of the worst-case bounds have in common that some background knowledge of the strength of shifts or confounding is needed to form and interpret these bounds. 
In contrast, we consider a model that shifts the distribution randomly. This allows to consider average distributional robustness. In addition, it
turns out that in such a random perturbation model, it is possible to estimate the size of perturbations by using knowledge in form of moment equations. Such background knowledge can come in the form of having multiple valid estimators for a single target quantity. Based on these estimators, it is possible to form "calibrated" confidence intervals that are valid on average, where we average both over sampling uncertainty and the distributional perturbation. 
Procedures to sample from the distributional perturbation model and conduct calibrated inference are implemented in the R-package calinf available at \url{github.com/rothenhaeusler/calinf}. 

Looking ahead, there are multiple directions that we believe are promising avenues for future research.

When having access to \emph{multiple datasets} or \emph{multi-source data}, we can model the different datasets arising from perturbed data generating distributions. In such a context, we point to the following. 

\paragraph*{Transfer learning under random shifts.}
In the literature, one often makes a covariate shift assumption, i.e.\ that the conditional distribution of a target $Y$ given a subset of observed attributes stays the same. The distributional perturbation model from Section~\ref{sec:calinf} allows to go beyond this assumption, by allowing for (non)-adversarial shifts even in conditional distributions. 
We can then formalize optimal transfer learning under random perturbations.

\paragraph*{Data fusion across heterogeneous datasets.} 
We can model the differences between multiple datasets as random, as in Section~\ref{sec:calinf}. This may lead to straightforward extensions of statistical methodology and optimality results (such as the Cram\'er-Rao lower bound or semi-parametric efficiency bounds) to distributional counterparts.

\paragraph*{Multiple testing in the context of distribution shifts.}
It is well known how to account for multiple testing in the context of sampling uncertainty. Similar issues are at play under multiple distributional perturbations: if we have 100 studies for which the null hypothesis holds but in each of those studies we sample from a randomly perturbed distribution $P' \neq P_{target}$, 
it is quite likely that we will get too many
false positives since some of the distributions will be strongly perturbed. The more perturbed distributions we look at, the more likely it is that we'll make a false discovery. This suggests that we should account for multiple testing also in distributional stability measures.

Without relying on multiple data sources, we also mention the following.
\paragraph*{Non-adversarial confounding.} Sensitivity analysis in causal inference investigates the stability of a causal conclusion by taking the worst-case confounded distribution given some restrictions on the strength of confounding. Such bounds are often very conservative. A less pessimistic assumption would be to model unobserved confounding as random (non-adversarial). A random confounding model, perhaps similar to the one in Section~\ref{sec:calinf}, potentially opens the door for novel average sensitivity procedures for causal inference.

\section*{Acknowledgments}
We thank the Guest Editors and the Editor for the opportunity of presenting
our work and the reviewers for constructive comments. The research of D. Rothenh\"ausler was supported by the Stanford Institute for Human-Centered Artificial Intelligence (HAI). The research was partially conducted during D. Rothenh\"ausler's research stay at the Institute for Mathematical Research at ETH Z\"urich (FIM). The research of P. B\"{u}hlmann was supported by the European Research Council under the Grant Agreement No 786461 (CausalStats - ERC-2017-ADG). 

\bibliographystyle{apalike}
\bibliography{references}

\end{document}